\documentclass[12pt,preprint]{aastex}
\slugcomment{Accepted for publication in ApJ}

\shorttitle{HCN(J = 3--2) and HCO$^{+}$(J = 3--2) lines in the circumnuclear region of NGC 1097}
\shortauthors{Hsieh et al.}

\begin{document}

\title{Probing Circumnuclear Environments with the HCN(J = 3--2) and HCO$^{+}$(J = 3--2) lines:  Case of NGC 1097}

\author{
        Pei-Ying Hsieh\altaffilmark{1,2},
        Paul T. P. Ho\altaffilmark{1,3},
        Kotaro Kohno\altaffilmark{4,6},
        Chorng-Yuan Hwang\altaffilmark{2},
        Satoki Matsushita\altaffilmark{1,5}
%et al.
%        Kohno Kotaro\altaffilmark{3},
%        Paul P. T. Ho\altaffilmark{2,4},
%        Satoki Matsushita\altaffilmark{2,5}
%        
%
\\pyhsieh@asiaa.sinica.edu.tw}

\affil{$^1$ Academia Sinica Institute of Astronomy and
       Astrophysics, P.O. Box 23-141, Taipei 10617, Taiwan, R.O.C.}
\affil{$^2$ Institute of Astronomy, National Central University,
        No.300, Jhongda Rd., Jhongli City, Taoyuan County 32001, Taiwan, R.O.C. }
\affil{$^3$ Harvard-Smithsonian Center for Astrophysics, 60 Garden Street, Cambridge, MA 02138, USA}
\affil{$^4$ Institute of Astronomy, The University of Tokyo, 2-21-1 Osawa, Mitaka-shi, Tokyo 181-0015}
\affil{$^5$ Joint ALMA Office, Alonso de C$\acute{\rm o}$rdova 3107, Vitacura 763 0355, Santiago, Chile}
\affil{$^6$ Research Center for the Early Universe, University of Tokyo, 7-3-1 Hongo, Bunkyo, Tokyo 113-0033, Japan}

\begin{abstract}

We present the first interferometric HCN(J = 3--2) and
HCO$^{+}$(J = 3--2) maps in the circumnuclear
region of NGC 1097, obtained with the Submillimeter Array.
The goal is to study the characteristics of the dense gas
associated with the starburst ring/Seyfert nucleus.
With these transitions, we suppress the diffuse low density emission in the nuclear region.  We detect and resolve the individual compact giant molecular
cloud associations (GMAs) 
in the 1.4 kpc circumnuclear starburst ring and within the
350 pc nuclear region.  
The nucleus is brighter than the ring in both lines,
and contributes to $\sim20\%$ and $\sim30\%$ to
the total detected HCO$^{+}$(J = 3--2) and HCN(J = 3--2) flux,
within the central 1.4 kpc.
The intensity ratios of
HCN(J = 3--2)/HCO$^{+}$(J = 3--2)
are roughly unity in the GMAs of the starburst ring. However,
this ratio is up to $\sim2$ in the nuclear region. From the
HCN(J = 3--2)/HCN(J = 1--0) ratio of $\le0.2$ in the nucleus,
we infer that the nuclear HCN(J = 3--2) emission might be optically
thin. The HCO$^{+}$(J = 3--2)
and HCN(J = 3--2) show correlations with $^{12}$CO(J = 3--2)
and the 24$\micron$~emission.
The tight correlations of HCN(J = 3--2), HCO$^{+}$(J = 3--2)
and 24$\micron$ emission in the starburst ring suggest that the
dense molecular gas and the dust are from the
same origins of star forming regions.
On the other hand, the HCN(J = 3--2) emission of
the nucleus is significantly enhanced, indicating mechanisms
other than star formation, such as AGN activities.
A self-consistent check of the fractional abundance
enhanced by X-ray ionization chemistry of the nucleus is possible
with our observations.
\end{abstract}


\section{INTRODUCTION}\label{sect-intro}

HCN and HCO$^{+}$ molecules
are important dense gas tracers, which are readily detectable in external   galaxies \citep[e.g.,][]{rickard77,nguyen89,jackson93,hen94,pag95,kohno03, gao04,riechers06,garcia10}, and in our own Galactic Center \citep[e.g.,][]{maria}.
With their larger dipole moments, these rotational lines have higher critical densities than that of CO and hence are excellent probes of the dense cores within the molecular clouds.  However, as the fractional abundance of HCN and HCO$^{+}$ are usually 1000 times less than CO, these lines are correspondingly fainter.  Nevertheless, extragalactic HCN emission lines conform to the well-known correlation with infrared emission over 7--8 orders of magnitude in sizescales \citep{gao04,wu05}. The HCN--IR luminosity has a tighter correlation than that of CO--IR \citep{sol92,sol97}.  This has been attributed to the fact that star formation occurs in dense gas rather than the diffuse gas traced by CO. However, in the extreme conditions of active galactic nuclei (AGN) and starbursts, HCN can be excited by the IR radiation rather than collisions \citep{aalto95,kohno08,sakamoto10,garcia10}, and the abundances of the HCN and HCO$^{+}$ can be affected by the X-ray irradiation \citep{tacconi94,lepp96,usero04,meijerink05, meijerink07,iman07,kohno08,harada10,garcia10}. Early multi-J statistical studies of the extragalactic HCN and HCO$^{+}$ were made by \citet{krips08} to study the molecular gas properties in AGNs and starbursts.
The abundance and excitation of these molecules are indeed affected by the circumnuclear environments.  Here, we focus on the J = 3--2 transitions of HCN and HCO$^{+}$, which have $\sim$30 times larger dipole moments as compared to CO, and have transition energy that are 6 times higher ($\sim$26 K) as compared to the J = 1--0 transitions.  Our purpose is to define the utility of these lines in the study of the dense and warm circumnuclear regions.

NGC 1097 is at a distance of 14.5 Mpc, and has a low luminosity-Seyfert nucleus and a circumnuclear starburst ring. The nucleus is prominent in X-ray, and is associated with a supermassive ($\sim$10$^{8}$M$_{\odot}$) black hole
\citep{lewis06,nemmen06}. The circumnuclear starburst ring shows prominent continuum emission at cm, mid-IR, and far-IR wavelengths, with compact knots, which are associated with star formation at a SFR of 5 M$_{\odot}$ yr$^{-1}$
\citep{hummel87,sandstrom10}. The general properties of the circumnuclear molecular gas of CO, HCN, and HCO$^{+}$ were studied with high resolution interferometers \citep{kohno01,kohno07,hsieh08,hsieh11}. These lower-J studies show
a molecular ring and a nuclear concentration, which are coincident with the starburst ring and the Seyfert nucleus. The HCN(J = 1--0) enhancement in the nucleus relative to the ring, has been interpreted as an X-ray irradiation effect on the HCN abundance \citep{kohno03,kohno07}. However, since the low-J lines only trace the cold and diffuse gas, higher-J lines may be better probes for the highly excited nuclear environments \citep[e.g.,][]{maria,matsu04}.  Here, we compare our first interferometric HCN(J = 3--2) and HCO$^{+}$(J = 3--2) maps of NGC 1097, at an angular resolution of $\sim$4\arcsec, with the emission from the lower-J transitions. 

\section{OBSERVATIONS AND DATA REDUCTION}\label{sect-obs}

We observed NGC 1097 with the Submillimeter Array\footnote{
The SMA is a joint pro ject between the Smithsonian
Astrophysical Observatory and the Academia Sinica Institute 
of Astronomy and Astrophysics and is funded by the
Smithsonian Institution and the Academia Sinica.} \citep{ho04}.
Two nights of observations were obtained in 2008 and 2010
with the compact configuration and the 345 GHz
receivers. The $uv$-coverages were $\sim$7.5 -- 61.5 $k\lambda$
for both observations. The system temperatures ($T_{\rm sys}$)
were $\sim$150--300 K and
$\sim$100--180 K, respectively, for the two sessions.

The SMA correlator processes two IF sidebands separated by
10 GHz, with $\sim$2 GHz bandwidth each in 2008.  The bandwidth
was upgraded to 4 GHz bandwidth for each IF after 2009.
The upper/lower sidebands were divided into slightly overlapping 24 and
48 chunks of 104 MHz width in 2008 and 2010, respectively.
In 2008, we observed the HCN(J = 3--2) line (rest frequency:265.886 GHz) in the LSB with 6 antennas.
In 2010, we observed both the HCN(J = 3--2) and the HCO$^{+}$(J = 3--2)
(rest frequency:267.558 GHz) lines simultaneously in the LSB with 8 antennas because of the availability of the 4 GHz bandwidth.

We calibrated the SMA data with the MIR-IDL software package.
The bandpass calibrators are
3C84, Uranus and Ceres for the first track, and
3C454.3 and 3C84 for the second track.
Callisto and Uranus were used for the absolute flux calibration
for the first and second dataset, respectively.
The accuracy of the absolute flux calibration is $\sim15$\%
for both of the data.
The gain calibrators
are the quasars 0423--013 for the first track and
0137--245 (phase)/0403-360 (amplitude) for the second track, respectively.
Mapping and analysis were done with the MIRIAD and the NRAO AIPS packages.
We made the maps with natural weighting, and with the data binned
to 20 km s$^{-1}$ velocity resolution. The CLEAN
processes were done in MIRIAD
to remove the sidelobes. The synthesized beams
are 3\farcs7$\times$2\farcs6 (PA = -10\degr) for the 
HCN(J = 3--2) maps and 4\farcs4$\times$2\farcs6 (PA = -15\degr)
for the HCO$^{+}$(J = 3--2) maps. For the latter comparisons
of both lines, we convolved the maps to 4\farcs4$\times$2\farcs7 (PA = --14.6\degr). The 1$\sigma$ noise levels are 11mJy beam$^{-1}$
for both lines.

\section{Results}

\subsection{High Excitation Dense Gas in the Circumnuclear Region of NGC 1097}\label{sect-result1}

\begin{center}
\begin{figure}
\epsscale{0.7}
\includegraphics[scale=0.25,angle=-90]{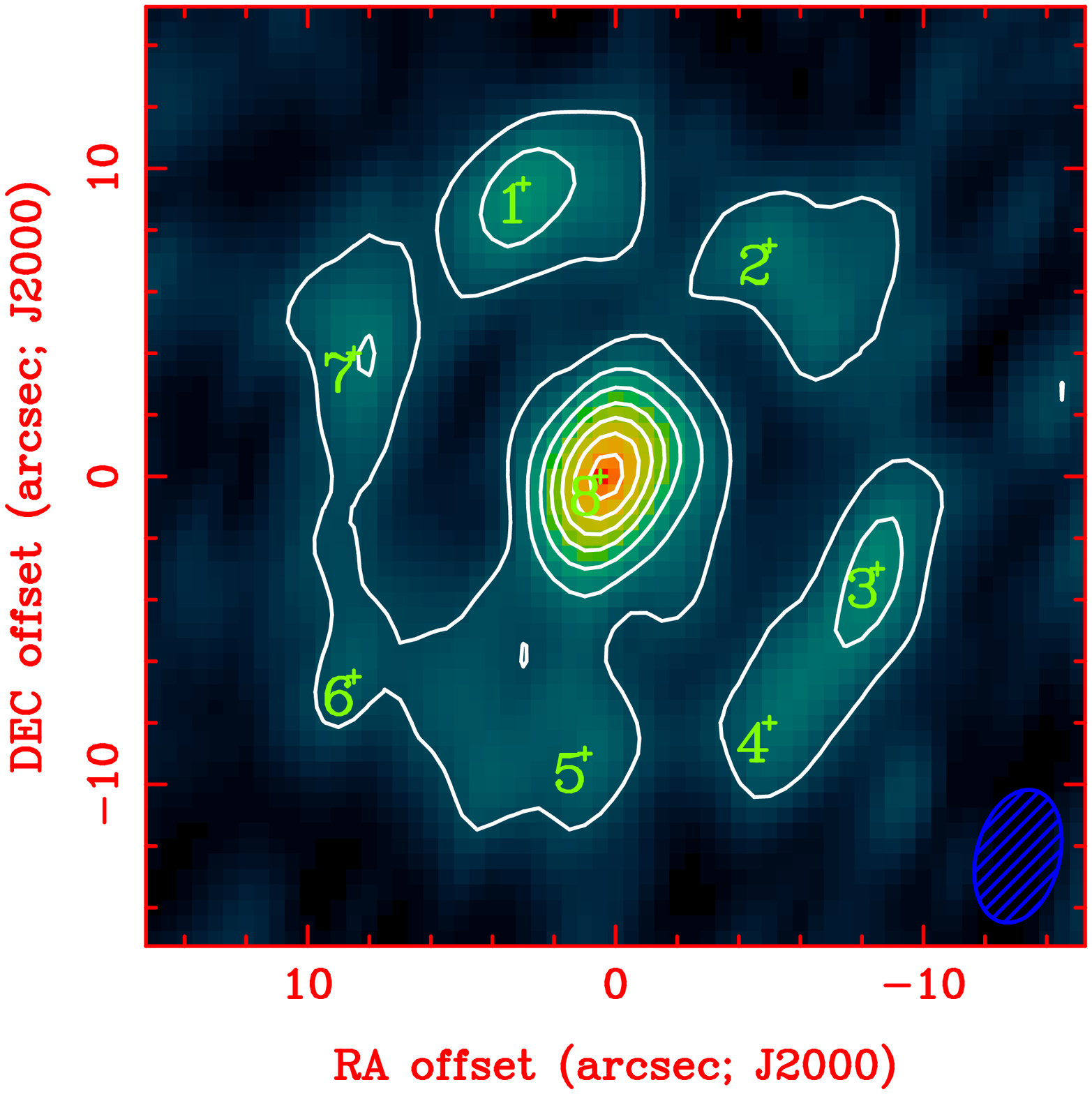}
\includegraphics[scale=0.25,angle=-90]{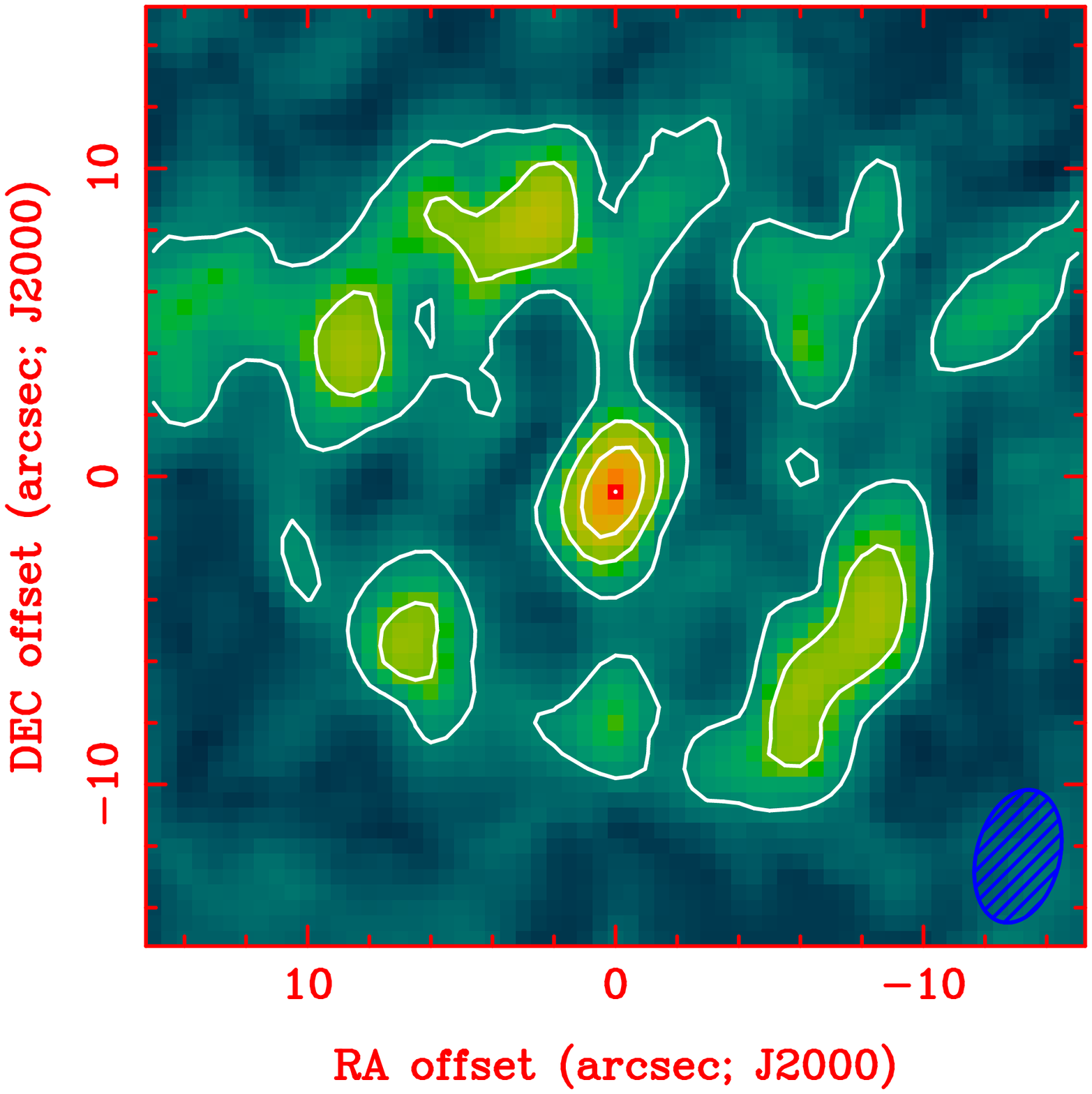}
\includegraphics[scale=0.25,angle=-90]{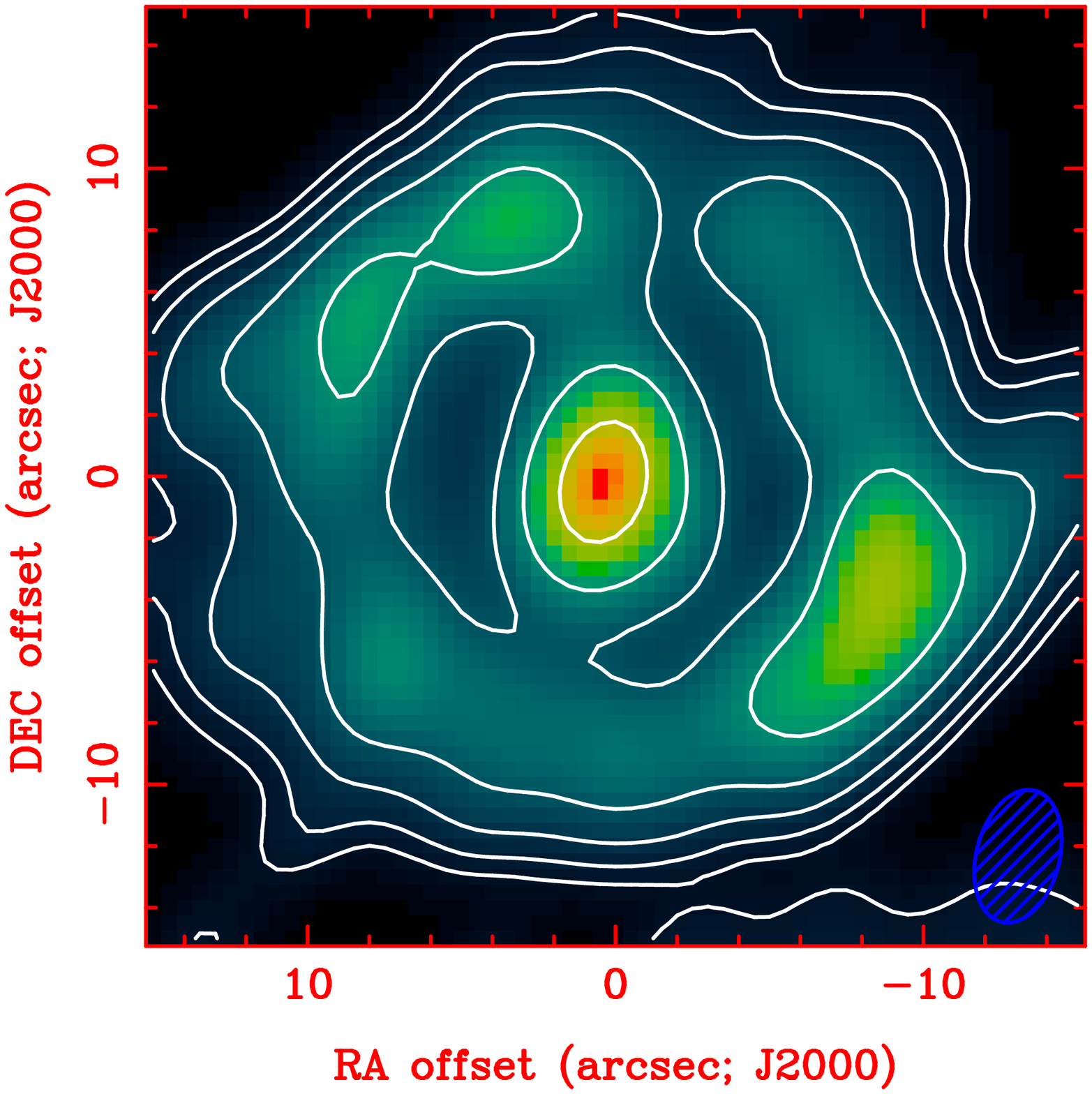}
\caption[]{Left: HCN(J = 3--2) integrated intensity map,
the contours levels are, 2, 4, 6, 8, 10, 12, 14$\sigma$, where 1$\sigma$ is 1.2 Jy beam$^{-1}$ km s$^{-1}$.
Middle:
HCO$^{+}$(J = 3--2) integrated intensity map,
the contours levels are, 2, 4, 6$\sigma$, where 1$\sigma$
is 1.2 Jy beam$^{-1}$ km s$^{-1}$.
Right: $^{12}$CO(J = 3--2) integrated intensity map,
the contours levels are, 2, 4, 8, 16, 32, 64$\sigma$, where 1$\sigma$ is 4.5 Jy beam$^{-1}$ km s$^{-1}$.
The synthesized beam of the three images are 4\farcs4$\times$2\farcs7
(PA = --14.6\degr).}
\label{fig-mom0}
\end{figure}
\end{center}

\begin{center}
\begin{figure}
\epsscale{0.7}
\includegraphics[scale=0.32,angle=0]{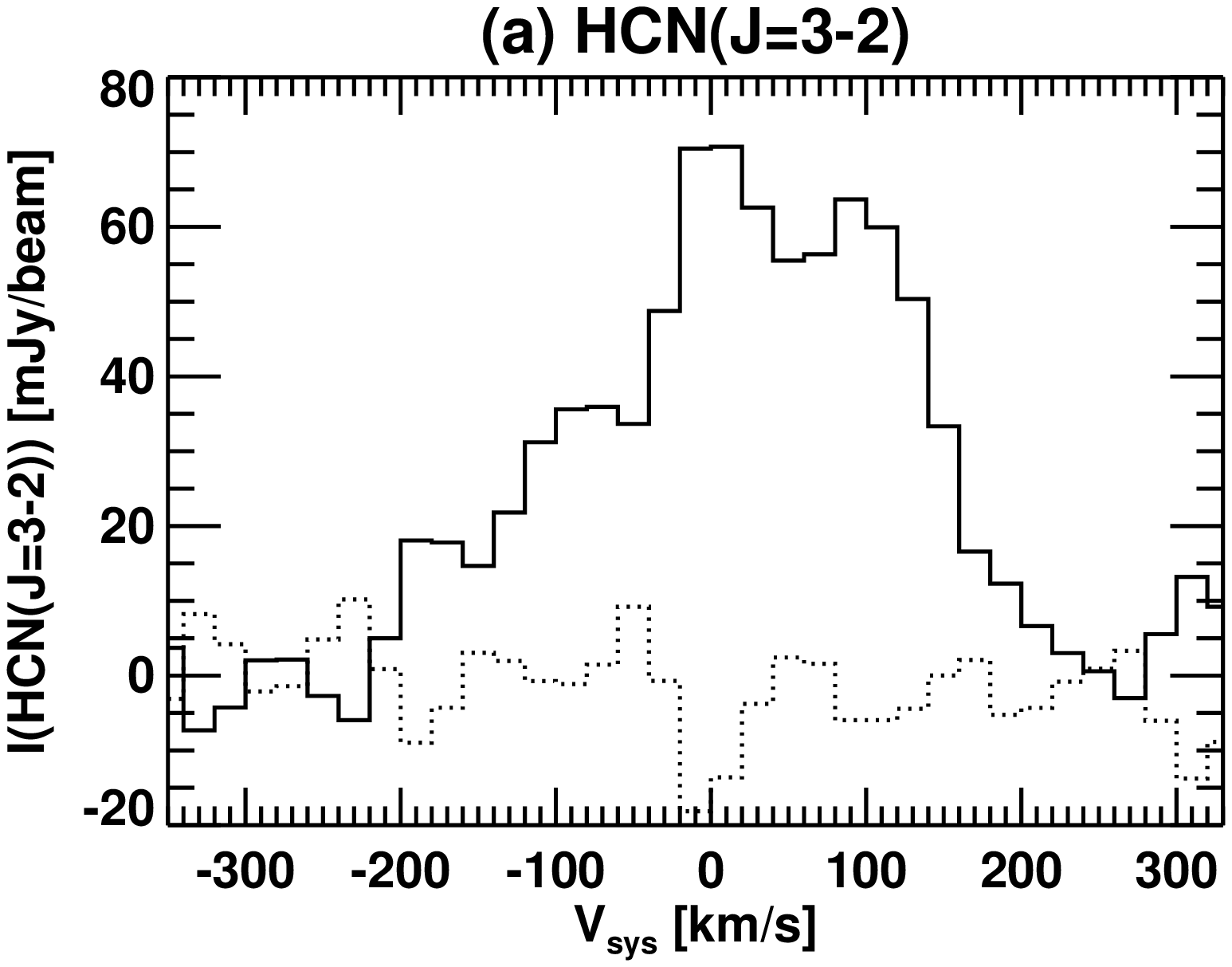}
\includegraphics[scale=0.32,angle=0]{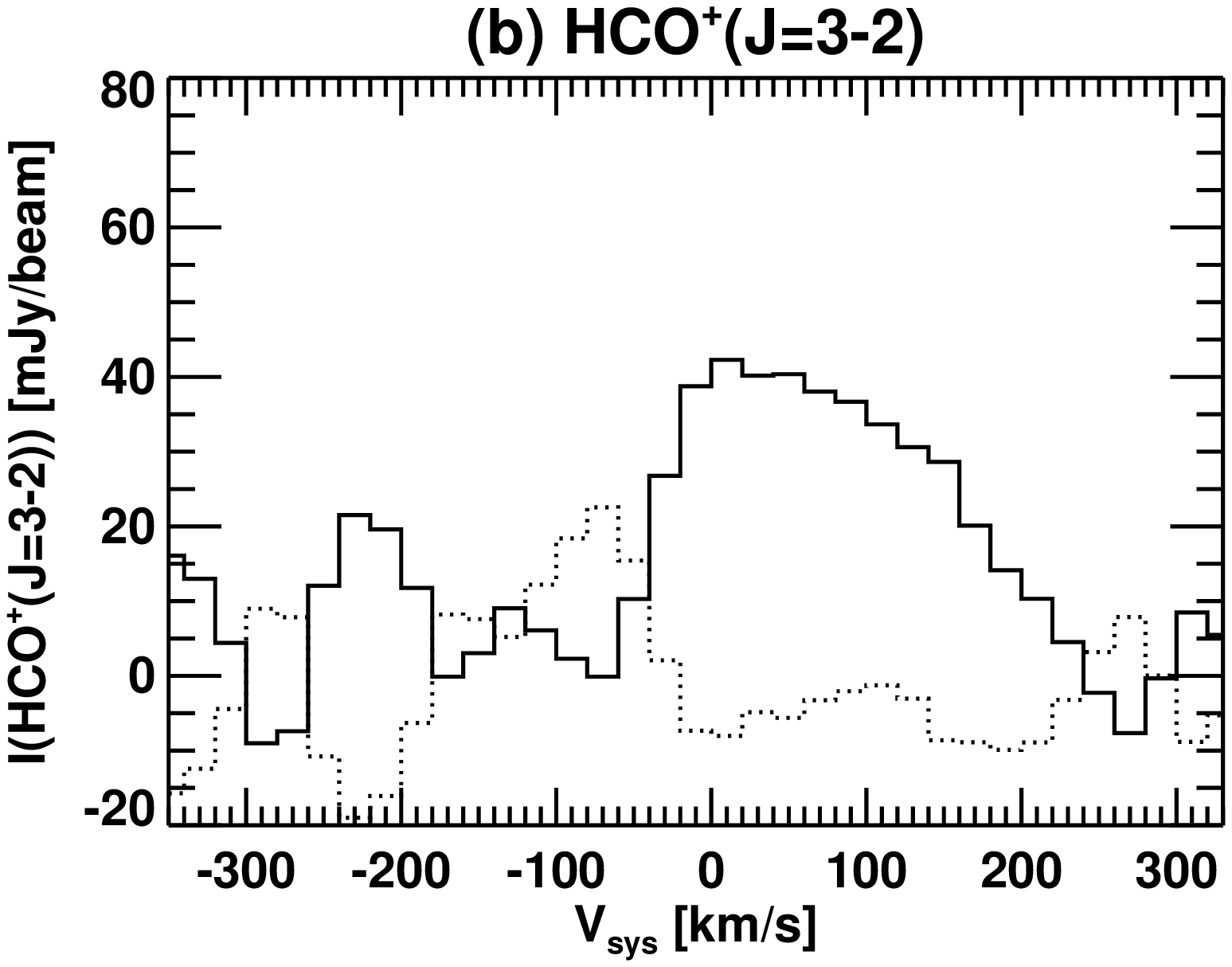}\includegraphics[scale=0.32,angle=0]{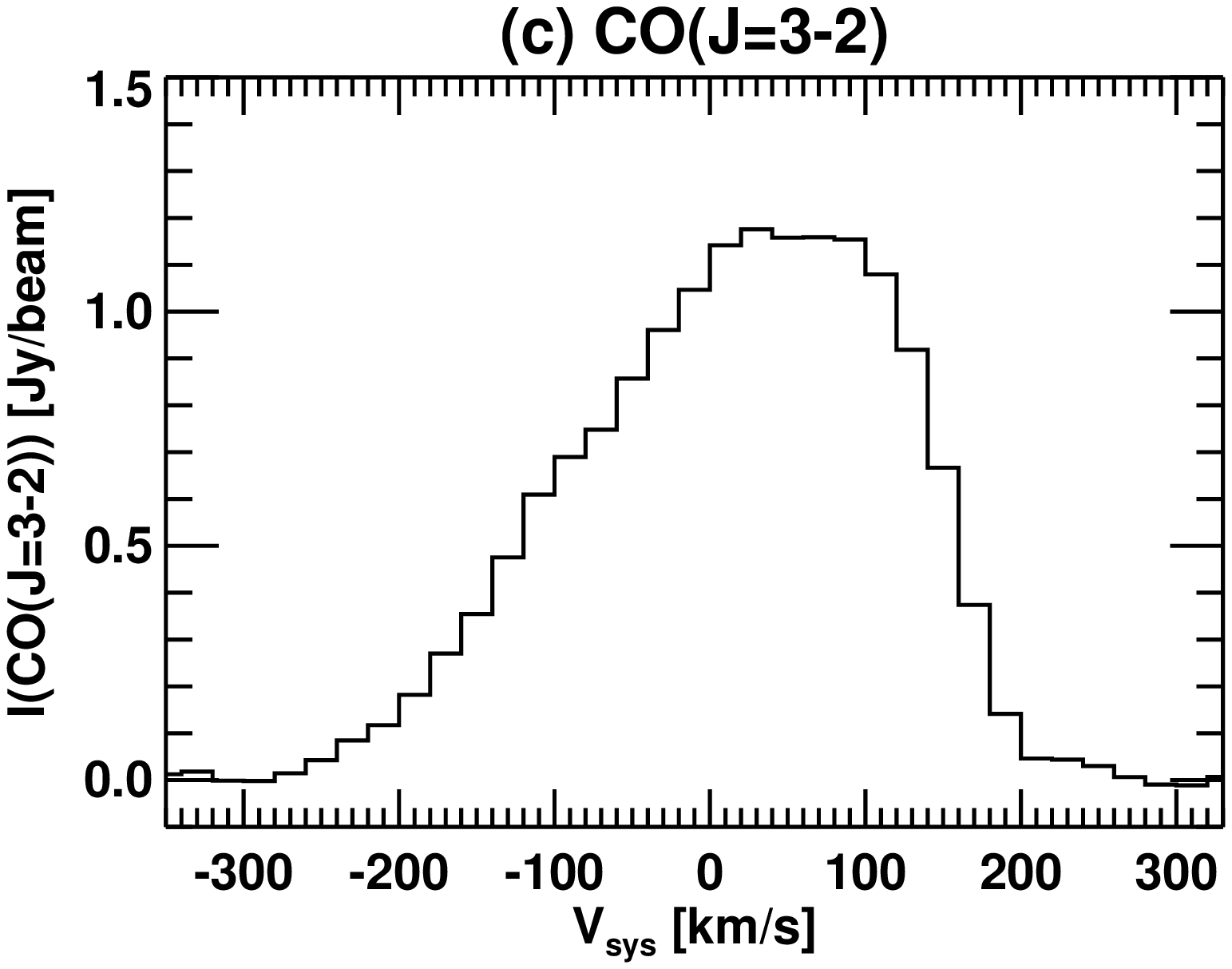}
%\plotone{12co21_chan_pa.ps}
\caption[]{
(a) HCN(J = 3--2) spectrum of the nucleus (solid-line).
We subtract the HCN(J = 3--2) spectrum by CO spectrum (Fig.2c) as template to check whether the residuals (dotted-line) are significant. The CO spectrum
is multiplied by 0.05 for the subtraction.
(b) HCO$^{+}$(J = 3--2) spectrum of the nucleus (solid-line).
We subtract the HCO$^{+}$(J = 3--2) spectrum by CO spectrum as template to check whether the residuals (dotted-line) are significant. The CO spectrum
is multiplied by 0.03 for the subtraction.
(c) CO(J = 3--2) spectrum of the nucleus.
The synthesized beam of all the spectra are 4\farcs4$\times$2\farcs7
(PA = --14.6\degr)}
\label{fig-spec}
\end{figure}
\end{center}

In Figure~\ref{fig-mom0} we show the HCN(J = 3--2),
HCO$^{+}$(J = 3--2) and $^{12}$CO(J = 3--2) \citep{hsieh11} integrated intensity maps. The HCN(J = 3--2) and HCO$^{+}$(J = 3--2) maps show a nuclear concentration and a molecular ring similar to the J = 1--0,
J = 2--1, and the J = 3--2 CO maps \citep{kohno03,kohno07,hsieh08,hsieh11}.
The nucleus is the brightest feature in all the maps, and the relatively weaker molecular ring coincides with the starburst ring. The ring consists of GMAs at the scale of giant molecular cloud associations (GMAs) of 200--300 pc. However, since both of the HCN and HCO$^{+}$ lines have higher critical density than CO, the HCN and HCO$^{+}$ GMAs are expected to be the denser regions of the GMAs mapped in the previous CO lines.
We identify a total of 8 GMAs including the one at the nucleus (Table~\ref{t.high.ratio}). The GMAs are defined by the intensity peaks with visual inspection.
We show the spectra of the nucleus in Figure~\ref{fig-spec}. The spectra show asymmetric profiles, which are likely due to multiple underlying components. The blueshifted part is fainter than the redshifted part in all the lines. 

The nucleus is brighter in HCN than HCO$^{+}$ in both the low- and high-J lines.  HCN(J = 3--2) has better contrast between the nucleus and the starburst ring than CO(J = 3--2) and HCO$^{+}$(J = 3--2). We demonstrate this by explicitly comparing the contribution of the emission from the nucleus and the molecular ring in these lines. The HCN(J = 3--2) integrated flux for the nucleus (radius: 0\arcsec--6\arcsec) and the ring (radius: 6\arcsec--13\arcsec) are 34.4$\pm$3.4 Jy km s$^{-1}$ and 89.9$\pm$6.7 Jy km s$^{-1}$, respectively. Hence the nucleus contributes 30$\pm3$\% of the HCN(J = 3--2) flux within 13\arcsec. The HCO$^{+}$(J = 3--2) integrated flux of the nucleus and the ring are 17.4$\pm$3.4 Jy km s$^{-1}$ and 85.8$\pm$6.7 Jy km s$^{-1}$, respectively.  Therefore, the nucleus contributes 17$\pm3$\% of the HCO$^{+}$(J = 3--2) flux within 13\arcsec. The integrated $^{12}$CO(J = 3--2) flux of the nucleus and ring are 1021 and 3778 Jy km s$^{-1}$.  Hence the nucleus contributes 21$\pm0.2$\% of the total $^{12}$CO(J = 3--2) flux within 13\arcsec.  The HCN(J = 3--2) is therefore significantly brighter in the nuclear region as compared with the molecular ring.

\subsection{Intensity Ratios of the Dense Gas in the Circumnuclear Region}\label{sect-result2}

We compare the HCN(J =3--2), HCO$^{+}$(J = 3--2), and $^{12}$CO(J = 3--2) \citep{hsieh11} at the angular resolution of 4\farcs4$\times$2\farcs7.  For the comparison of the J = 3--2 and J = 1--0 lines, we limited the shortest baselines to 7.5 $k\lambda$ for all transitions, in order to match the resolution of the  J = 1--0 lines at 7\farcs9$\times$3\farcs2. HCN(J =1--0), HCO$^{+}$(J = 1--0), and $^{12}$CO(J = 1--0) lines \citep{kohno07,kohno03,kohno11} are included for these comparisons.

\begin{center}
\begin{figure}
\epsscale{1}
\includegraphics[scale=1]{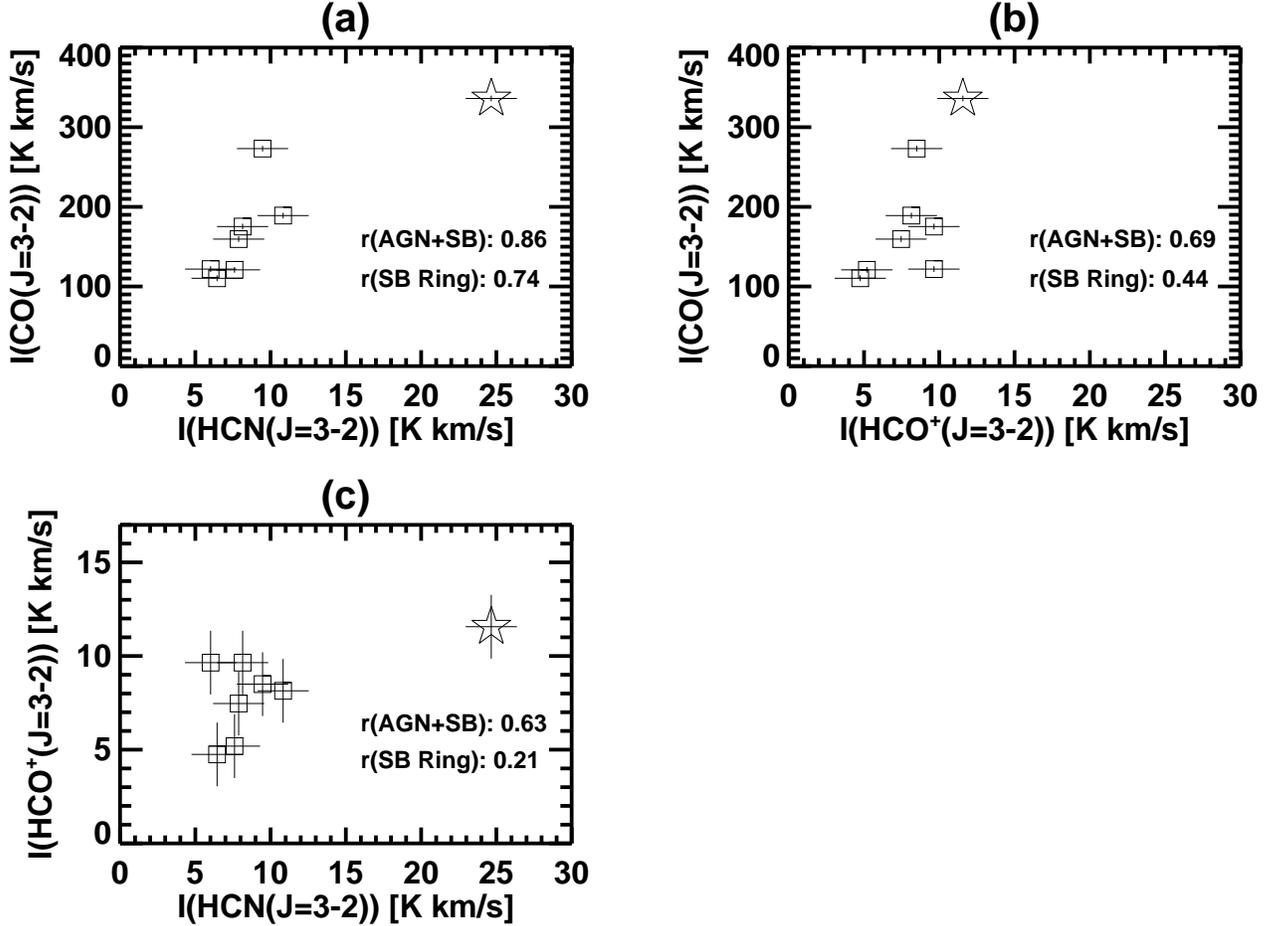}
%\plotone{12co21_chan_pa.ps}
\caption[]{(a) The correlation of HCN(J = 3--2) and $^{12}$CO(J = 3--2)
line intensity. The fluxes of the GMAs in the ring are shown
as squares, and star represents the data of the nucleus.
The units are K km s$^{-1}$.
The $\pm1\sigma$
uncertainties are overlaid on the symbols with vertical/horizontal bars.
The linear Pearson coefficients are shown in the plots. r(AGN+SB) are
the coefficients of all data, and r(SB Ring) are derived for the starburst ring
GMAs only.
(b) The correlation of HCO$^{+}$(J = 3--2) and $^{12}$CO(J = 3--2)
line intensity.
(c) The correlation of HCN(J = 3--2) and HCO$^{+}$(J = 3--2).
The symbols are the same as (a).
The data we show here are from resolution
of 4\farcs4$\times$2\farcs7.
}
\label{fig-hcn-high}
\end{figure}
\end{center}

In Table~\ref{t.high.ratio}, the averaged intensity ratios of HCN(J = 3--2)/HCO$^{+}$(J = 3--2) are 1.11$\pm$0.36 and 2.13$\pm$0.34 in the ring and the nucleus, respectively. These values are similar to the single dish observations compiled by \citet{krips08}, where the ratio is nearly unity in the starburst galaxies and larger than unity in the galaxies which host AGNs. The average intensity ratios of HCN(J = 3--2)/$^{12}$CO(J = 3--2) are 0.05$\pm$0.01 and 0.07$\pm$0.01 in the ring and the nucleus, respectively.
In Figure~\ref{fig-hcn-high} we show the correlation of the fluxes of HCN(J = 3--2), HCO$^{+}$(J = 3--2), and $^{12}$CO(J = 3--2) lines.
While there is a general correlation, the dispersion of these ratios within the molecular ring is about a factor of 2.
We calculated the linear Pearson correlation coefficients of the plots in Figure~\ref{fig-hcn-high}. The correlation coefficients of all data are 0.86, 0.69, and 0.63
for Figure~\ref{fig-hcn-high}(a), (b), and (c) respectively.
The correlation coefficients of the GMAs in the starburst ring are
0.74, 0.44, and 0.21 for Figure~\ref{fig-hcn-high}(a), (b), and (c).
The higher coefficients represent stronger linear dependence of the two quantities.
This variation in the molecular ring is responsible for the apparent differences between the GMAs in the ring.  Such differences are marginally significant but difficult to interpret.  Chemistry, excitation, and the contributions from diffuse gas to the 
CO(J = 3--2), may all be important.    

We compare the ratios of J = 3--2 and J = 1--0 lines in low resolution (7\farcs9$\times$3\farcs2) in Table~\ref{t.low.ratio}. \citet{kohno01} reported that the lower-J line ratios of HCN(J = 1--0)/$^{12}$CO(J = 1--0) is $\sim$0.34, and the HCN(J = 1--0)/HCO$^{+}$(J = 1--0) is $\sim$2 in the nucleus. The HCN/HCO$^{+}$ seems to be similar in lower- and higher-J lines, while the HCN/$^{12}$CO decreases toward the higher-J lines in the nucleus (0.04$\pm$0.01). The ring shows similar trends in Table~\ref{t.low.ratio}. In both nucleus and ring, the HCO$^{+}$/CO ratios decrease in J = 3--2.
On the other hand, the HCN(J = 3--2)/HCN(J = 1--0) and HCO$^{+}$(J = 3--2)/HCO$^{+}$(J = 1--0) ratios does not show significant difference in the ring and the nucleus.
This result is consistent with the survey done by \citet{krips08}. The decreasing HCN/CO ratios
toward higher--J lines were reported in other galaxies host AGN,
for instance, NGC 1068 \citep{krips08,krips11}, and M51 \citep{matsu98,matsu04,matsu07}. We will discuss the physical
conditions of the molecular gas in the Sect.~\ref{sect-discussion}.

\begin{center}
\begin{figure}
\epsscale{1}
\includegraphics[scale=1]{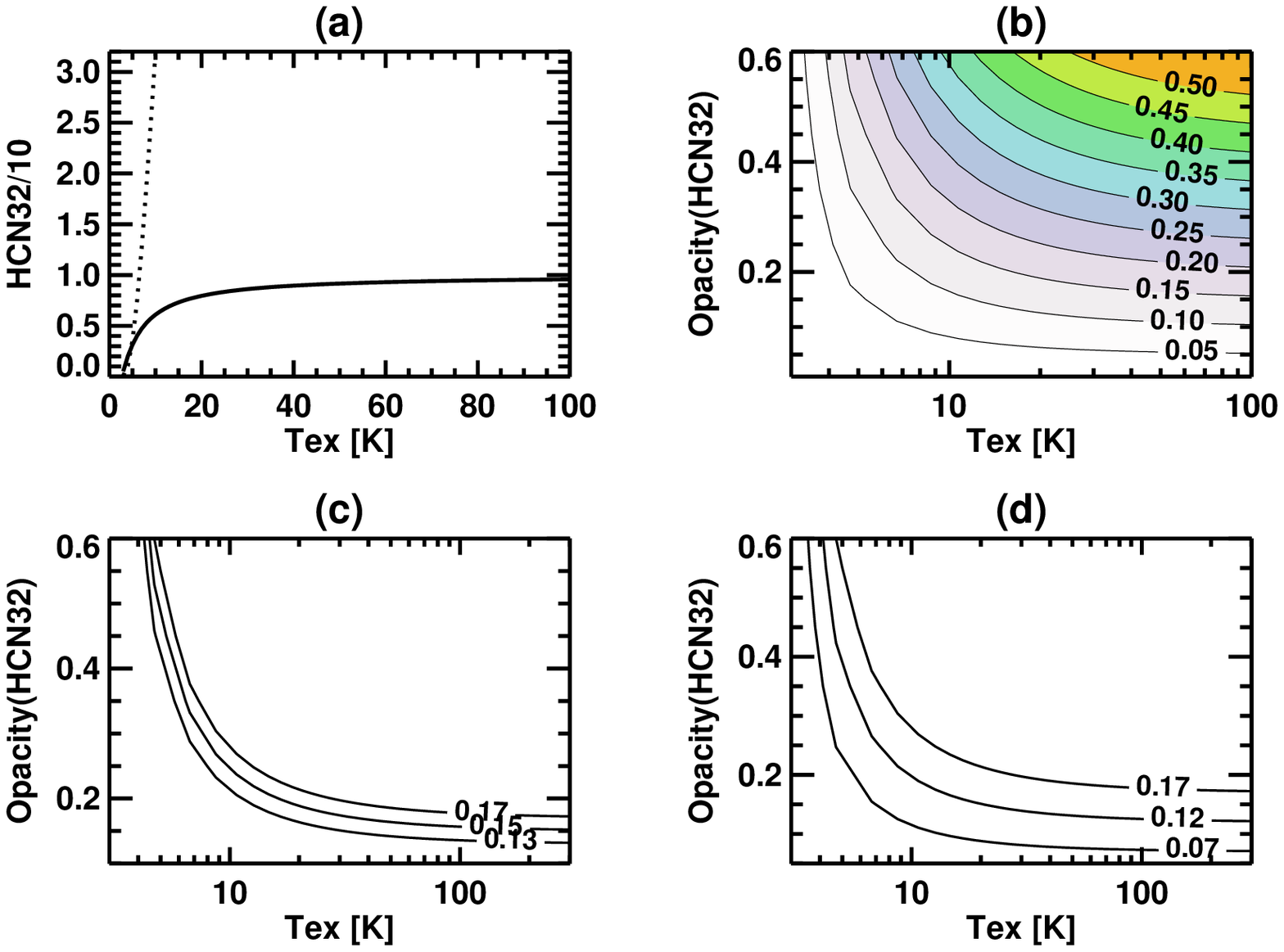}
%\plotone{12co21_chan_pa.ps}
\caption[]{(a) The HCN(J = 3--2)/HCN(J = 1--0) intensity ratio
as a function of excitation temperature in LTE conditions.
The solid line is for when both lines are optically thick. The
dotted line is for when both lines are optically thin.
(b) The HCN(J = 3--2)/HCN(J = 1--0) intensity ratio
as a function of excitation temperature and
$\tau_{\rm HCN32}$ in LTE conditions.
In this case, the HCN(J = 3--2) is thin and HCN(J = 1--0)
is thick. The ratios are shown as contours from 0.05 to 0.5
in steps of 0.05.
(c) The HCN(J = 3--2)/HCN(J = 1--0) ratios of the nucleus
0.15$\pm$0.02 shown as a function of the $T_{\rm ex}$
and $\tau_{\rm HCN_{32}}$.
(d) The HCN(J = 3--2)/HCN(J = 1--0) ratios of the ring
0.12$\pm$0.05 shown as a function of the $T_{\rm ex}$
and $\tau_{\rm HCN_{32}}$.}
\label{fig-lte-hcn}
\end{figure}
\end{center}

\begin{center}
\begin{figure}
\epsscale{1}
\includegraphics[scale=1]{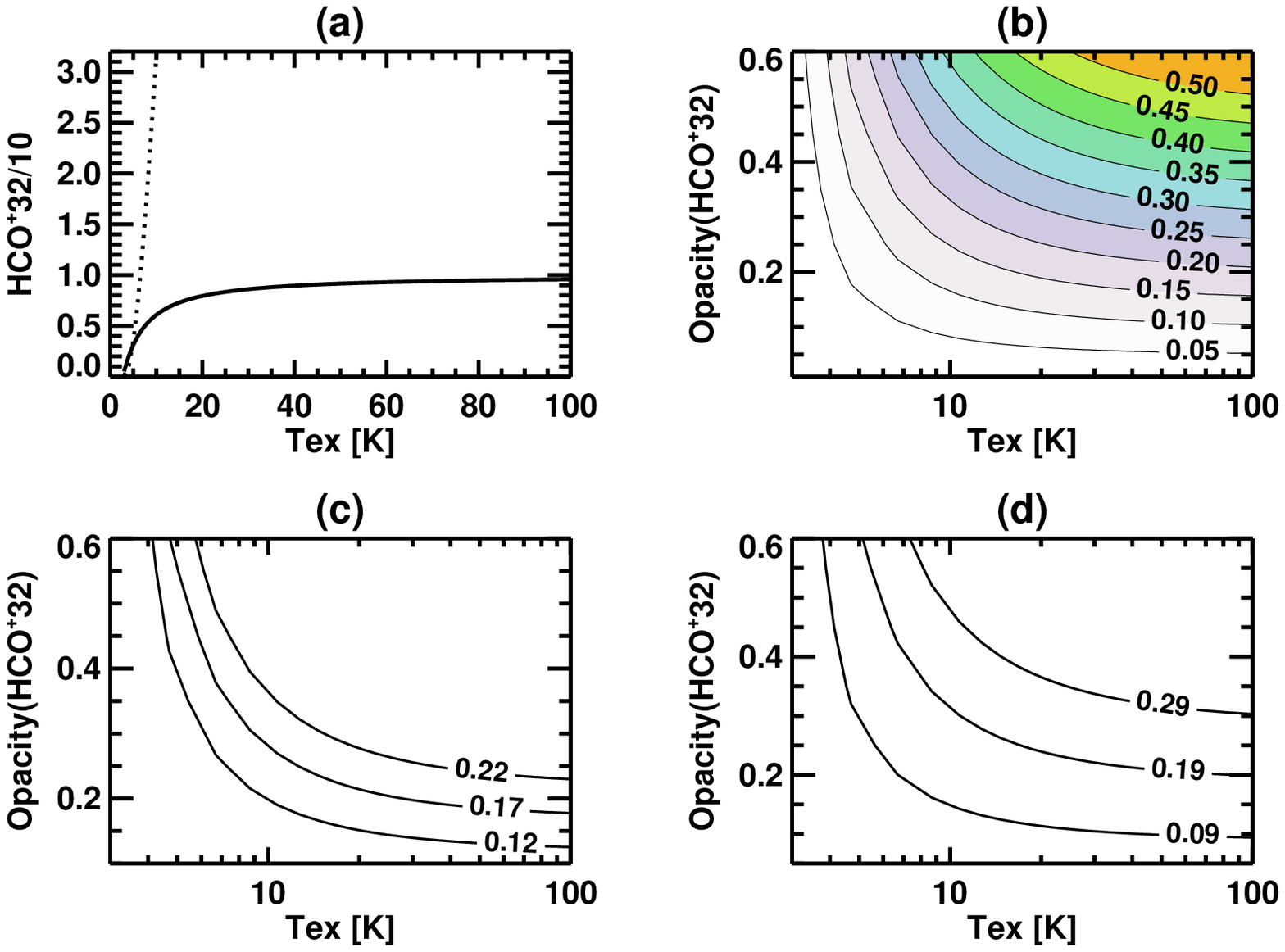}
\caption[]{(a) The HCO$^{+}$(J = 3--2)/HCO$^{+}$(J = 1--0) intensity ratio
as a function of excitation temperature in LTE conditions.
The solid line is for when both lines are optically thick. The
dotted line is for when both lines are optically thin.
(b) The HCO$^{+}$(J = 3--2)/HCO$^{+}$(J = 1--0) intensity ratio
as a function of excitation temperature and $\tau_{\rm HCO^{+}_{32}}$
in LTE conditions.
In this case, the HCO$^{+}$(J = 3--2) is thin and HCO$^{+}$(J = 1--0)
is thick. The ratios are shown as contours from 0.05 to 0.5
in steps of 0.05.
(c) The HCO$^{+}$(J = 3--2)/HCO$^{+}$(J = 1--0) ratios of the nucleus
0.17$\pm$0.05 shown as a function of the $T_{\rm ex}$
and $\tau_{\rm HCO^{+}_{32}}$. 
(d) The HCO$^{+}$(J = 3--2)/HCO$^{+}$(J = 1--0) ratios of the ring
0.19$\pm$0.10 shown as a function of the $T_{\rm ex}$
and $\tau_{\rm HCO^{+}_{32}}$. } 
\label{fig-lte-hco}
\end{figure}
\end{center}

\begin{center}
\begin{figure}
\epsscale{1}
\includegraphics[scale=1]{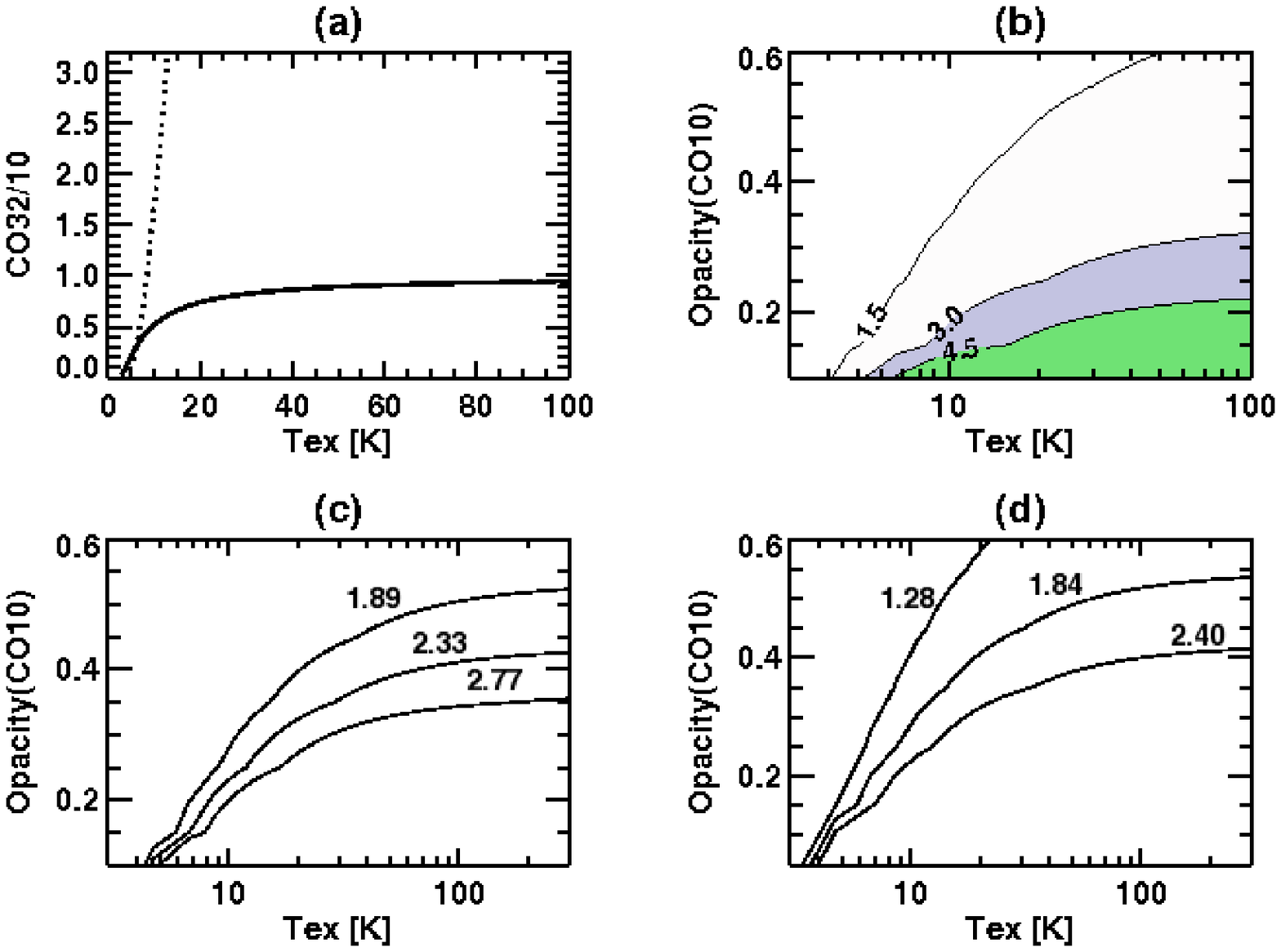}
\caption[]{(a) The $^{12}$CO(J = 3--2)/$^{12}$CO(J = 1--0) intensity ratio
as a function of excitation temperature in LTE conditions.
The solid line is for when both lines are optically thick. The
dotted line is for when both lines are optically thin.
(b) The $^{12}$CO(J = 3--2)/$^{12}$CO(J = 1--0) intensity ratio
as a function of excitation temperature and $\tau_{\rm CO_{10}}$
in LTE conditions.
In this case, the $^{12}$CO(J = 1--0) line is thin and $^{12}$CO(J = 3--2) line
is thick. The ratios are shown as contours higher than unity,
from 1.5, 3, and 4.5 to illustrate the dependence of the opacity
and $T_{\rm ex}$.
(c) The $^{12}$CO(J = 3--2)/$^{12}$CO(J = 1--0) ratios of the nucleus
2.33$\pm$0.44 shown as a function of the $T_{\rm ex}$
and $\tau_{\rm CO_{10}}$.
(d) The $^{12}$CO(J = 3--2)/$^{12}$CO(J = 1--0) ratios of the ring
1.84$\pm$0.56 shown as a function of the $T_{\rm ex}$
and $\tau_{\rm CO_{10}}$.} 
\label{fig-lte-co}
\end{figure}
\end{center}

We derive the line ratios in LTE conditions to roughly
look into the effects of temperatures, opacities and abundance
ratios on the HCN and HCO$^{+}$ lines.
A more sophisticated radiative transfer model will be important
once we have more lines and higher angular resolution.
In our case, we are looking at the large scale properties of the molecular gas,
and a simple LTE analysis is adequate since all the information
are averaged by the coarse resolution.
We define the brightness temperature
$T_{\rm B}$ = $f[J_{\nu}(T_{\rm ex})-J_{\nu}(T_{\rm bg})](1-e^{-\tau})$,
where $T_{\rm B}$ is the brightness temperature, $f$ is the filling
factor, $J_\nu$ = $(h\nu/k)(1/e^{h\nu/kT-1})$, $T_{\rm ex}$ is
the excitation temperature, $T_{\rm bg}$ is the background
temperature, and $\tau$ is the opacity.
In our estimates we assume the filling factors are the same
for both lines. For the same molecular
species at J = 3--2 and J = 1--0, the estimates are:

\begin{enumerate}
\setlength{\tabcolsep}{0.02in}
\item{
Both lines are optically thick ($\tau \gg 1$), in which case the
integrated intensity line ratio can be written as
\begin{equation}
R_{32/10}^{\rm thick} = {[J_{\nu_{32}}(T_{\rm ex}) - J_{\nu_{32}}(T_{\rm bg})]
\over{[J_{\nu_{10}}(T_{\rm ex}) - J_{\nu_{10}}(T_{\rm bg})] }} \label{thickratio},
\end{equation}
where the background temperature $T_{\rm bg}$ is 2.7K.
In LTE, the $R_{32/10}$ approaches to unity with rising temperatures (see Figure~\ref{fig-lte-hcn}a and Figure~\ref{fig-lte-hco}a).}
\item 
Both lines are optically thin ($\tau \ll 1$) and the integrated
intensity line ratio can be written as 
\begin{equation}
R_{32/10}^{\rm thin} = {[J_{\nu_{32}}(T_{\rm ex}) - J_{\nu_{32}}(T_{\rm bg})]\tau_{32}
\over{[J_{\nu_{10}}(T_{\rm ex}) - J_{\nu_{10}}(T_{\rm bg})]\tau_{10} }},\label{thinratio}
\end{equation}
and
\begin{equation}
\frac{\tau_{32}}{\tau_{10}} = 3\frac{\nu_{32}}{\nu_{10}}e^{-\frac{h\nu_{21}}{kT_{\rm ex}}}e^{-\frac{h\nu_{10}}{kT_{\rm ex}}}\frac{1-e^{-h\nu_{32}/kT_{\rm ex}}}{1-e^{-h\nu_{10}/kT_{\rm ex}}},\label{tauratio}
\end{equation}
which in LTE depends on the optical depth $\tau$ at each frequency.
$\tau_{32}$ and $\tau_{10}$ are the opacity of J = 3--2
and J = 1--0 lines. In Figures~\ref{fig-lte-hcn}a and ~\ref{fig-lte-hco}a, the ratio
increases rapidly with $T_{\rm ex}$.
\item{
We considered J = 3--2 line is optically thin and J = 1--0 line is optically thick
\begin{equation}
R_{32/10}^{\rm thin/thick} = {[J_{\nu_{32}}(T_{\rm ex}) - J_{\nu_{32}}(T_{\rm bg})]
\over{[J_{\nu_{10}}(T_{\rm ex}) - J_{\nu_{\rm 10}}(T_{\rm bg})]}}\tau_{32}. \label{thinratio}
\end{equation}}
In Figure~\ref{fig-lte-hcn}b and Figure~\ref{fig-lte-hco}b, we
illustrate the dependence of the ratios on temperature
and opacity, where the ratios of the same molecules are
independent of the abundance. In both figures, the ratios
seem to be more sensitive to the temperatures when $T_{\rm ex}\le$
20K, but are dependent on the opacity of HCN(J = 3--2) and
HCO$^{+}$(J = 3--2) at $T_{\rm ex}\ge$ 20K. This dependence
of temperature/opacity also varies with ratios.

\end{enumerate}
\setlength{\tabcolsep}{0.02in}
The ratios of $^{12}$CO(J = 3--2)/$^{12}$CO(J = 1--0)
lines are shown in Table~\ref{t.low.ratio}.
We also derive the LTE results of the
$^{12}$CO(J = 3--2)/$^{12}$CO(J = 1--0) ratios in Figure~\ref{fig-lte-co}.
The first two cases in Figure~\ref{fig-lte-co}a are for when both lines are optically thick and
optically thin. The third case (Figure~\ref{fig-lte-co}b) is assuming
optically thin $^{12}$CO(J = 1--0) and optically thick $^{12}$CO(J = 3--2) lines.

\begin{center}
\begin{figure}
\epsscale{1}
\includegraphics[scale=1]{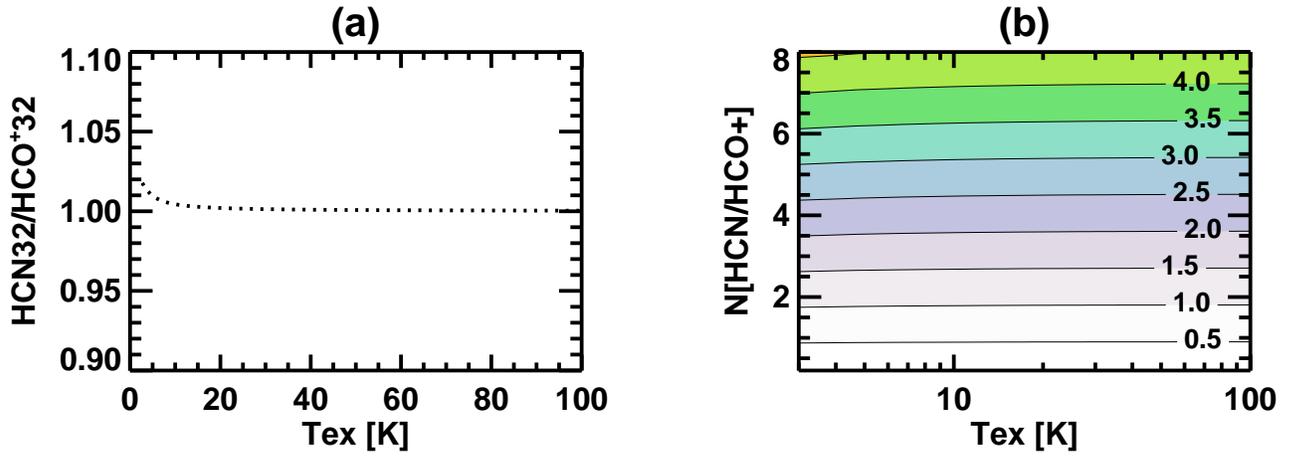}
\caption[]{(a) The HCN(J = 3--2)/HCO$^{+}$(J = 3--2) intensity ratios
as a function of excitation temperature in LTE conditions.
In this case both lines are optically thick. The ratios are
nearly unity.
(b) The HCN(J = 3--2)/HCO$^{+}$(J = 3--2) intensity ratios (contours)
as a function of excitation temperature and abundance ratios
of HCN/HCO$^{+}$ in LTE conditions. The ratios of the
contours are labeled.
In this case, both lines are optically thin.}
\label{fig-lte-hcn-hco}
\end{figure}
\end{center}

For the ratios of the different species of HCN(J = 3--2) and HCO$^{+}$(J = 3--2),
we illustrate the two cases when both lines are optically thick and
thin:

\begin{enumerate}
\item{
Both lines are optically thick ($\tau \gg 1$), in which case the
integrated intensity line ratio can be written as
\begin{equation}
R_{\rm HCN/HCO^{+}}^{\rm thick} = {[J_{HCN}(T_{\rm ex}) - J_{HCN}(T_{\rm bg})]
\over[J_{HCO^{+}}(T_{\rm ex}) - J_{HCO^{+}}(T_{\rm bg})]}. \label{thickratio}
\end{equation}
The ratios quickly saturate to nearly unity (see Figure~\ref{fig-lte-hcn-hco}a and Figure~\ref{fig-lte-hcn-hco}a) in the case when both lines are thick.}
\item{
Both lines are optically thin ($\tau \ll 1$) and the integrated
intensity line ratio can be written as 
\begin{equation}
R_{\rm HCN/HCO^+}^{\rm thin} = {[J_{HCN}(T_{\rm ex}) - J_{HCN}(T_{\rm bg})]\tau_{HCN}
\over{[J_{HCO^{+}}(T_{\rm ex}) - J_{HCO^{+}}(T_{\rm bg})]\tau_{HCO^{+}} }},\label{thinratio}
\end{equation}
and
\begin{equation}
\footnotesize
\frac{\tau_{\rm HCN}}{\tau_{\rm HCO^{+}}} = \frac{\nu^{2}_{\rm HCO^{+}32}}{\nu^{2}_{\rm HCN32}}
\frac{N_{\rm HCN}}{N_{\rm HCO^{+}}}
\frac{B_{\rm HCN32}}{B_{\rm HCO^{+}32}}
\frac{e^\frac{\rm -6hB_{HCN32}}{kT_{\rm ex}}}{e^\frac{\rm -6hB_{HCO^{+}32}}{kT_{\rm ex}}}
\frac{A_{\rm HCN32}}{A_{\rm HCO^{+}32}}
\frac{1-e^{\rm -h\nu_{HCN}/kT_{\rm ex}}}{1-e^{\rm -h\nu_{HCO^{+}}/kT_{\rm ex}}},
\label{tauratio}
\end{equation}
which in LTE depends on the optical depth $\tau$ at each frequency.
$\tau_{\rm HCN}$ and $\tau_{\rm HCO^{+}}$ represent the opacity of HCN(J = 3--2)
and HCO$^{+}$(J = 3--2) lines. The $B_{\rm HCN32}$ and $B_{\rm HCO^{+}32}$
are the rotational constants of HCN and HCO$^{+}$.
$A_{\rm HCN32}$ and $A_{\rm HCO^{+}32}$ are the Einstein A coefficients
of HCN(J = 3--2) and HCO$^{+}$(J = 3--2). $N_{\rm HCN}$
and $N_{\rm HCO^{+}}$ are the total column density
of HCN and HCO$^{+}$ molecules. Therefore the free
parameters are the column density ratio, which is proportional
to the abundance ratio of the HCN and HCO$^{+}$ molecules if
their size scales
and the excitation temperature $T_{\rm ex}$ are similar.
In Figure~\ref{fig-lte-hcn-hco}b,
we show the HCN(J = 3--2)/HCO$^{+}$(J = 3--2) ratios
as a function of $T_{\rm ex}$ and HCN/HCO$^{+}$ abundance ratio.
The line ratios are nearly insensitive to the temperatures, but are
dependent on the abundance ratios.}

\end{enumerate}

\subsection{Correlation of the molecular line intensity and 24$\micron$ continuum}
	
\begin{center}
\begin{figure}
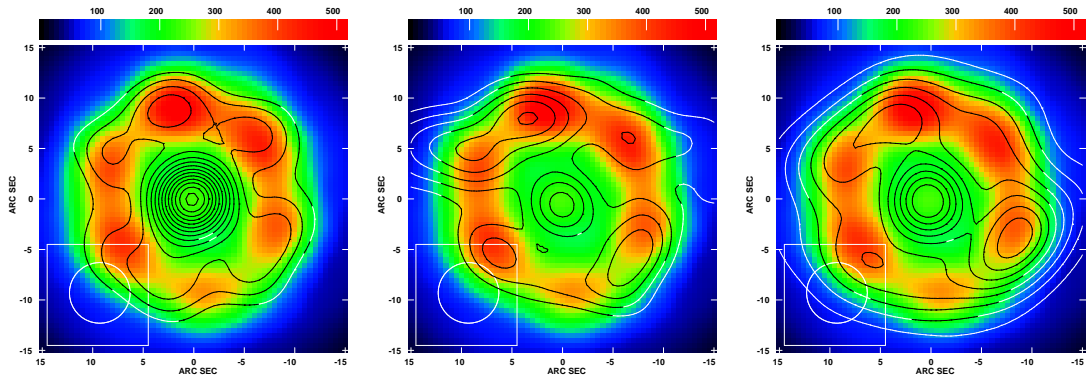

\epsscale{0.7}
\includegraphics[scale=0.25]{fig7a.ps}
\includegraphics[scale=0.25]{fig7b.ps}
\includegraphics[scale=0.25]{fig7c.ps}
\caption[]{Left: The HCN(J = 3--2) map is overlaid
on the 24$\micron$ continuum map. Middle: The HCO$^{+}$(J = 3--2)
map is overlaid on the 24$\micron$ continuum  map.
Right: The CO(J = 3--2) map is overlaid on the 24$\micron$ continuum map.
The resolutions of the HCN(J = 3--2) and HCO$^{+}$(J = 3--2)
are matched to that of the PSF of the 24$\micron$ continuum map (6\arcsec).
The contour starts from 4.50 Jy Beam$^{-1}$ km s$^{-1}$, and
in steps of 1.50 Jy Beam$^{-1}$ km s$^{-1}$.}
\label{fig-24}
\end{figure}
\end{center}

\begin{center}
\begin{figure}
\epsscale{1}
\includegraphics[scale=1]{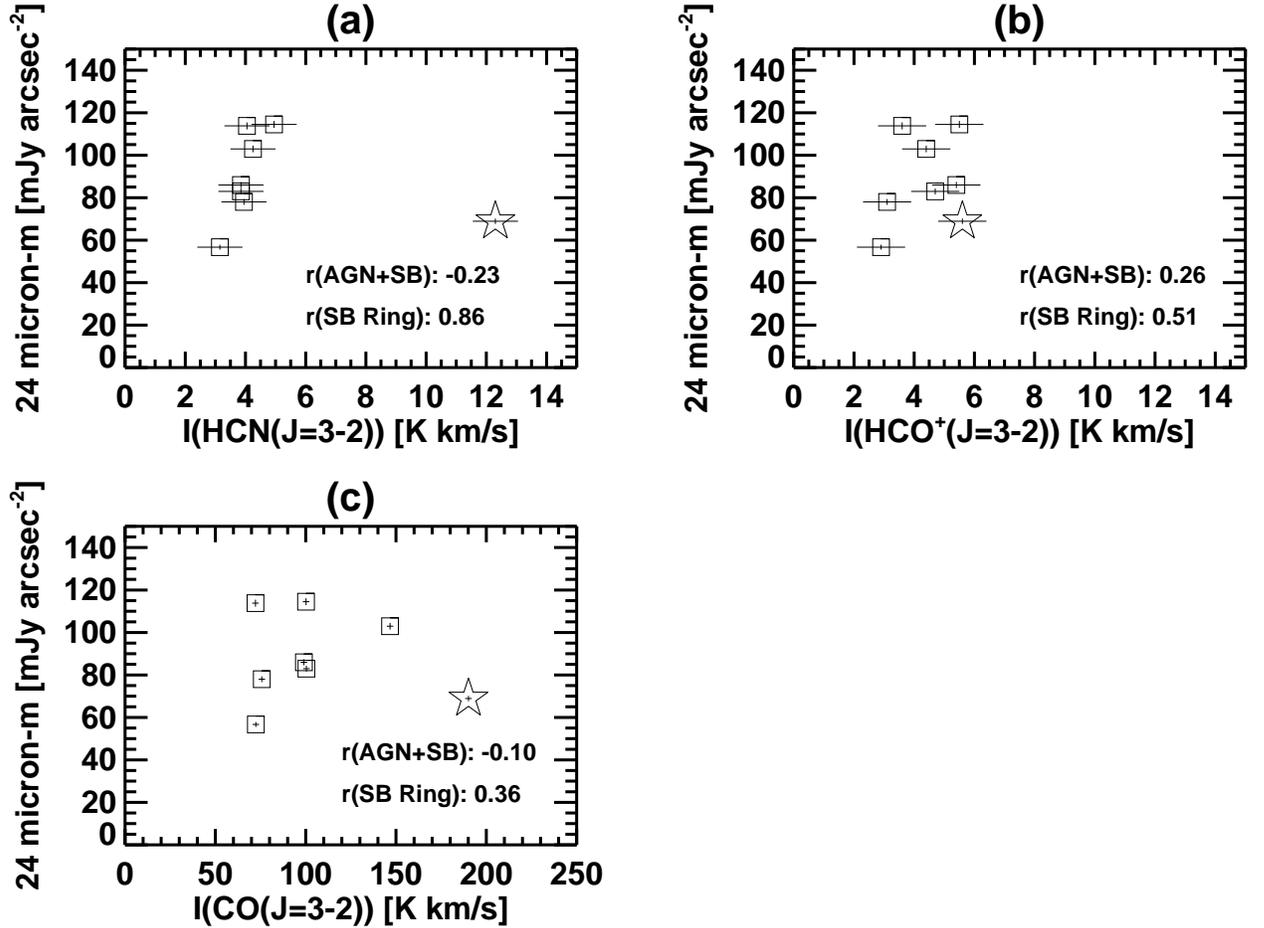}
\caption[]{(a) The correlation of HCN(J = 3--2) integrated intensity
and 24$\micron$ continuum flux. The symbols are the same as
Figure~\ref{fig-hcn-high}. The $\pm1\sigma$
uncertainties are overlaid on the symbols with vertical/horizontal bars.
The linear Pearson coefficients are shown in the plots. r(AGN+SB) are
the coefficients of all data, and r(SB Ring) are derived for the starburst ring
GMAs only.
(b) The correlation of HCO$^{+}$(J = 3--2) integrated intensity
and 24$\micron$ continuum flux. The symbols are the same as
Figure~\ref{fig-hcn-high}. 
(c) The correlation of $^{12}$CO(J = 3--2) integrated intensity
and 24$\micron$ continuum flux. The symbols are the same as
Figure~\ref{fig-hcn-high}. The flux are measured from
low resolution maps (6\arcsec). 
}
\label{fig-hcn-low}
\end{figure}
\end{center}

In Figure~\ref{fig-24} we show the comparison of the molecular gas and the archival Spitzer 24$\micron$ continuum emission. The PSF of the 24$\micron$ map is $\sim$6\arcsec, and we smoothed our molecular line maps to match this resolution.
In Figure~\ref{fig-hcn-low} we show the relation of the molecular line intensity with the 24$\micron$ continuum emission for the GMAs. In Figure~\ref{fig-hcn-low}(a), the HCN(J = 3--2) flux of the ring seems to be tightly correlated to the 24$\micron$ continuum emission, while the nuclear emission is offset from this correlation. The relation is less tight for the lower critical density lines of HCO$^{+}$(J = 3--2) and $^{12}$CO(J = 3--2). However, for these lines, the nuclear emission is closer to the correlation of the ring.
We calculated the linear Pearson correlation coefficients of the plots in Figure~\ref{fig-hcn-low}.
The correlation coefficients of all data are --0.23, 0.26, and --0.10
for Figure~\ref{fig-hcn-low}(a), (b), and (c) respectively.
The correlation coefficients of the GMAs in the starburst ring are
0.86, 0.51, and 0.36 for Figure~\ref{fig-hcn-low}(a), (b), and (c).
In this case, including the nucleus will decrease the linear dependence of the two quantities.
	
\section{DISCUSSIONS}\label{sect-discussion}

The nucleus of NGC 1097 seems to have environments of higher
excitation than the ring.
Our spatially resolved HCN(J = 3--2) and HCO$^{+}$(J = 3--2) maps
show that the nucleus is brighter than the molecular ring.
Similar results have also been shown in other interferometric
studies of multi-J HCN/HCO$^{+}$ emission such
as M51 and NGC 1068 \citep{matsu98,matsu04,matsu07,krips11}.
The ratio of upper transition as compared to lower transitions, on the other hand, does not show a sharp enhancement in the nuclear region.  The nature of the higher excitation in the nucleus can thus be deduced as following discussions. The necessity of analysis with multiple transitions has already been indicated from the previous studies by \citet{kohno03} and the single dish survey by \citet{krips08} toward nearby active/starburst galaxies. In this paper, we discuss our
results with standard LTE analysis.

\subsection{Physical Properties of the J = 3--2 lines}

In Sect.~\ref{sect-result2} we derived the LTE estimates
of the physical parameters for the HCN and HCO$^{+}$ molecules.
In Figures~\ref{fig-lte-hcn}a and \ref{fig-lte-hco}a, if the HCN(J = 3--2),
HCN(J = 1--0), HCO$^{+}$(J = 3--2), and HCO$^{+}$(J = 1--0)
lines are optically thick,
which suggests cold/thick dense
molecular gas in both the nucleus and the ring with $T_{\rm ex}\sim5$K.
This is possible since our spatial resolution is still quite coarse at
300 pc, so that we are likely sampling extended and colder material. 
However, since the optically thick molecular gas has typical temperature of $\ge$10K \citep{dishoeck}, while the observed HCN(J = 3--2)/HCN(J = 1--0)
and HCO$^{+}$(J = 3--2)/HCO$^{+}$(J = 1--0) ratios are lower
than LTE estimates of $\sim0.5$ at 10K. The inconsistence
may be either the four lines are not all optically thick, or the different
filling factors of those transitions.

In Figure~\ref{fig-lte-hcn-hco}a, the HCN$_{32}$/HCO$^{+}_{32}$
ratio is nearly unity if both lines are thick. This is consistent
with the average ratio of the ring, but inconsistent
for the nucleus. Therefore, both HCN(J = 3--2) and HCO$^{+}$(J = 3--2)
being optically thick is less likely for the nucleus.
This case can not be ruled out since we do not have a promising
optically thin line to determine the physical parameters.
Alternatively, that the J = 3--2 lines are weaker than the J = 1--0
lines, may be because the J = 3--2 lines are thin and J = 1--0 lines
are thick. As it is shown in Figures~\ref{fig-lte-hcn}b and \ref{fig-lte-hco}b,
the line ratios depend on the J = 3--2 opacity and $T_{\rm ex}$.
Note that the J = 3--2 to J = 1--0 ratios
for the nucleus and the ring, are similar within the uncertainties.
The LTE estimates suggest that the J = 3--2 lines
may be colder with higher opacity or warmer with lower opacity.

In the case where both HCN(J = 3--2) and HCO$^{+}$(J = 3--2)
are optically thin, we could estimate the abundance ratio
of HCN/HCO$^{+}$ in Figure~\ref{fig-lte-hcn-hco}b.
The [HCN]/[HCO$^{+}$] ratio is $\sim2.5-4.5$ and $\sim1-3$
for the nucleus and the ring.
The optically thin scenarios
seem to be supported by Figure~\ref{fig-hcn-low}a,b,
which show tighter correlations with the 24$\micron$ emission
at GMAs scale. The tighter relations of HCN(J = 3--2),
and HCO$^{+}$(J = 3--2) with 24$\micron$ emissions
suggest these two lines trace the star forming regions
better than CO(J = 3--2) line.
If the HCN(J = 3--2), HCO$^{+}$(J = 3--2), and 24$\micron$ emissions
are optically thin, their fluxes will be proportional to the column density of
molecular gas and dust. Their good
correlations may reflect that these are reliable tracers of the mass.
On the other hand, the abundant molecule of CO is likely to remain optically thick
in the J = 3--2 line, which leads to a poor relation in
Figure~\ref{fig-hcn-low}c.
We note that the enhancement of the HCN(J = 3--2) emission in the nucleus
is not simple. The HCN(J = 3--2)
has a factor of two excess flux as compared to the HCO$^{+}$(J = 3--2).
This excess is likely due to the different abundance ratio
of the HCN and HCO$^{+}$ in the ring and the nucleus
as it is shown in Figure~\ref{fig-lte-hcn-hco}b. The abundance
ratio would depend on the different physical environments,
and we will discuss the effects of the chemistry in the following section.

On the contrary to the low HCN(J = 3--2)/HCO$^{+}$(J = 3--2) ratios,
the high $^{12}$CO(J = 3--2)/$^{12}$CO(J = 1--0) ratios in the
circumnuclear region of NGC 1097
seem to suggest that the $^{12}$CO(J = 3--2) line is optically thick,
and the $^{12}$CO(J = 1--0) is optically thin (Figure~\ref{fig-lte-co}).
The similar high ratios $\ge$2 were also reported in NGC 1068 \citep{krips11,tsai11} and M51 \citep{matsu04}. In Figure~\ref{fig-lte-co},
for the case of $^{12}$CO(J = 3--2)/$^{12}$CO(J = 1--0) ratios
larger than unity, the higher $T_{\rm ex}$ or lower opacity
of $^{12}$CO(J = 1--0) line lead to the higher
$^{12}$CO(J = 3--2)/$^{12}$CO(J = 1--0) ratios.
Given the similar $^{12}$CO(J = 1--0) opacity
of the nucleus and the molecular ring, the higher
$^{12}$CO(J = 3--2)/$^{12}$CO(J = 1--0) ratio of the nucleus
suggests it has higher temperature than the ring, or
the nucleus has lower $^{12}$CO(J = 1--0) opacity than the ring.
These factors might also be consistent with the decreasing
HCN/CO ratios toward high--J lines reported in the previous
studies by \citet{krips08,krips11}.
Note that the LTE analysis gives the simplified results since the ratios
also depend on other physical parameters such as gas density.
Unfortunately, the LTE analysis can not provide straightforward
density solution. This is because we need
additional information such as fractional abundance and
source size. However, in principle we expect in the case of collisional
excitation, higher density conditions might
increases the higher--J population of molecules.

In terms of excitation, the observed low values for the ratios of HCN(J = 3--2)/HCN(J = 1--0) and HCO$^{+}$(J = 3--2)/HCO$^{+}$(J = 1--0) also suggest that the J = 3--2 transitions might be subthermal.  The case of subthermal excitation for the higher-J HCN/HCO$^{+}$ lines has also been found in other galaxies \citep[e.g.,][]{nguyen92,riechers11}.  The critical density is defined when the molecules are in thermal equilibrium via collisions. Densities have to be higher in order to drive the collisional time scale to be significantly shorter than the spontaneous decay time scale as defined by the Einstein A coefficient, in order to populate the upper energy levels.  Since the Einstein A coefficient scales as $\nu$$^{3}$, the rapid spontaneous decay rate will eventually overwhelm the collision rate for a sufficiently high energy state.  In such case, the subthermally excited lines will set an upper limit on the densities as compared to
the critical density of the particular transitions. Based on the bright HCN/HCO$^{+}$ J = 3--2 emission, and their low intensity ratio as compared to the lower transitions, the low density scenario with subthermal excitation, is also possible.
However, the high $^{12}$CO(J = 3--2)/$^{12}$CO(J = 1--0) ratios
of the circumnuclear region in NGC 1097 might suggest high gas
density, which is in conflict to the above scenario.
A careful excitation analysis cooperating with multi-J molecular lines
are essential to derive the gas density, excitation temperature, and
kinetic temperature to examine the possibility of subthermal
conditions of J = 3--2 HCN, HCO$^{+}$ lines, such as the works
done by \citet{krips08,krips11}.

\subsection{Column Density of the HCN and HCO$^{+}$ molecules}\label{sect-discussion2}

We can calculate the column density of HCN and HCO$^{+}$ by assuming LTE conditions and correcting for the optical depth,  

\begin{equation}
N_{mol}=\frac{3k}{8\pi^{3}B_{rot}\mu^{2}}\frac{e^{hB_{rot}J_{l}(J_{l}+1)/kT_{ex}}}{J_{l}	+1}\frac{(T_{ex}+hB_{rot}/3k)}{(h\nu/k)e^{-h\nu/kT_{ex}}}
\frac{\tau_{\nu}}{1-e^{-\tau_{\nu}}}\int{T_{b} dv},
\end{equation}
where $N_{\rm mol}$ is the column density of specific molecule,  $k$ and $h$ are the Boltzmann and Planck constants, $B_{\rm rot}$ is the rotational constant of the molecule, $\mu$ is the electric dipole moment of the molecule, $J_{l}$ is the quantum number of the lower energy level from $J+1\rightarrow J$, $T_{\rm ex}$ is the excitation temperature which is constant under LTE assumption for all transitions, $\tau_{\nu}$ is the opacity, and $dv$ is the velocity width. The column density of H$_{2}$ can then be derived from the abundance ratio of [Molecule]/[H$_{2}$].  Alternatively, if we derive the column density of H$_{2}$ from other observations, we can estimate the abundances of HCN and HCO$^{+}$. We calculate $B_{\rm rot}$ of HCN and HCO$^{+}$ to be 44.31 and 44.59 GHz.  The values of $\mu$ for HCN and HCO$^{+}$ are 2.98$\times10^{-18}$ and 3.92$\times10^{-18}$ e.s.u. cm. We then calculate $N_{\rm HCN}$ and $N_{\rm HCO^{+}}$ by
	
\begin{equation}
N_{\rm HCN}=1.108\times10^{10}\frac{(T_{ex}+0.709)}{e^{-25.525/T_{ex}}}\frac{\tau_{\nu}}{1-e^{-\tau_{\nu}}}\int{T_{b}(HCN(J=3-2)) dv},
\end{equation}
	
\begin{equation}
N_{\rm HCO^{+}}=6.325\times10^{9}\frac{(T_{ex}+0.713)}{e^{-25.685/T_{ex}}}\frac{\tau_{\nu}}{1-e^{-\tau_{v}}}\int{T_{b}(HCO^{+}(J=3-2)) dv},
\end{equation}.

We estimate the cases of the nucleus and the ring separately with
different temperatures and opacities.
As for the nucleus, we adopt the warm and thin conditions based
on the previous discussions.
In terms of temperatures, \citet{sandstrom10} fitted the SED of the nucleus of NGC 1097 for the spatial scale of 600 pc, and found that the thermal blackbody spectrum peaks at $\sim100\micron$, corresponding to a dust temperature of $T_{\rm d}\sim30$K. However, an
additional peak seems to be located at
$\sim$10$\micron$, where corresponds to $T_{\rm d}$
of $\sim300$K. We thus constrain the range of the $N_{\rm HCN}$
within this range. Since the opacities are nearly constant
from 30 to 300 K (Figure~\ref{fig-lte-hcn}b), 
the corresponding
opacities of HCN(J = 3--2) and HCO$^{+}$(J = 3--2) are $\sim$0.2
in this range. Then $N_{\rm HCN}$ is $\sim$2$\times10^{13}$ to 1$\times10^{14}$ cm$^{-2}$, and
$N_{\rm HCO^+}$ is $\sim$6$\times10^{12}$ to 3$\times10^{13}$
cm$^{-2}$.
The column density of H$_{2}$ in the nucleus, as derived from CO(J = 3--2) (Table~\ref{t.high.ratio}) with a conversion factor of 1.8$\times10^{20}$ (K km s$^{-1}$)$^{-1}$ cm$^{-2}$ \citep{dame}, is 3$\times10^{22}$ cm$^{-2}$.
If we adopt this value for the total H$_{2}$ column density, we can estimate
the fractional abundance of HCN and HCO$^{+}$. We find $N_{\rm HCN}$/$N_{\rm H_2}$ is 7$\times10^{-10}$ to 3$\times10^{-9}$, and
$N_{\rm HCO^+}$/$N_{\rm H_2}$ is 2$\times10^{-10}$ to 9$\times10^{-10}$.
The conversion factor in the galactic center was suggested to be
lower than the value of galactic disk. \citet{mauer} suggests that
the conversion factor is
$0.3\times10^{20}$ (K km s$^{-1}$)$^{-1}$ cm$^{-2}$ in the nuclear starburst environments. Hence the abundances mentioned above would be
larger by a factor of 6.
The key uncertainty here is whether the CO, HCN, and HCO$^{+}$, are spatially coextensive. The derived column density ratios are the mean values averaged over the volume of the CO emission. 
Another route to estimate the [HCN]/[H$_{2}$] is to adopt the
$N_{\rm H_2}$ as derived from the X-ray absorption. The
$N_{\rm H_2}$ of the nucleus as derived by \citet{nemmen06}
is 4.6$\times10^{20}$ cm$^{-2}$, which increases
the [HCN]/[H$_{2}$] and [HCO$^{+}$]/[H$_{2}$] by a factor of $\sim$60.
%to 4$\times10^{-8}$--2$\times10^{-7}$ and (1--6.5)$\times10^{-8}$, respectively.

As for the molecular star-forming ring, using the same
analysis, with a temperature of 30 K, and opacities
of 0.2, we derive $N_{\rm H_2}$ for the ring-GMAs of
3$\times10^{22}$ cm$^{-2}$, $N_{\rm HCN}$ of $\sim7\times10^{12}$ cm$^{-2}$,
and $N_{\rm HCO^{+}}$ of $\sim4\times10^{12}$ cm$^{-2}$.
The [HCN]/[H$_{2}$] and [HCO$^{+}$]/[H$_{2}$] are
2$\times10^{-10}$ and 1$\times10^{-10}$, respectively.
%If we adopt $N_{\rm H_2}$ for the ring as derived
%from the X-ray absorption of 1.5$\times10^{21}$ cm$^{-2}$,
%the [HCN]/[H$_{2}$] and [HCO$^{+}$]/[H$_{2}$] would be increased
%by a factor of $\sim$20.

CO is usually optically thick because of its large abundance and
low critical density. The concern is that we may sample
the cold gas on the surface of the molecular clouds with CO,
while the HCN and HCO$^{+}$ sample the inner region
of the clouds. The H$_{2}$ column density as
derived from the X-ray absorption may be more appropriate
if the X-ray traces the more compact and denser regions.
We thus favor the larger fractional abundance as
estimated above via the X-ray absorption.

\subsection{Chemical Abundances in the Nucleus}

We can check if these chemical abundances are consistent with the ionization chemistry as suggested in the classical studies of \citet{lepp96}.  These authors predicted the fractional abundance of molecules for various ionizing densities (see their Fig. 2, 3). The ionizing flux densities was defined by $\xi$/$n_{\rm H_{2}}$ (erg s$^{-1}$ cm$^{-3}$) in their paper, where $\xi$ and $n_{\rm H_{2}}$ are the ionization rate and the volume number density of the molecular hydrogen. If the ionizing flux is dominated by the X-ray emission in the nucleus, its luminosity within the energy range of 2--10 KeV is 4.4$\times10^{40}$ erg s$^{-1}$ \citep{nemmen06}. \citet{maloney96} derived the X-ray ionization rate in the low-ionization limit as $\xi\sim1.4\times10^{-11}L_{44}r_{2}^{-2}N_{22}^{-1}~(\rm s^{-1})$, where $L_{44},~r_{2},~N_{22}$ are the X-ray luminosity in unit of 10$^{44}$ erg s$^{-1}$, the distance to the X-ray source in the unit of 100 pc, and the H$_{2}$ column density attenuating the X-ray flux in the unit of 10$^{22}$ cm$^{-2}$.
We assume the ionization radiation has an isotropic distribution, and we adopt the $N_{\rm H_{2}}$ as derived from CO of 3$\times10^{22}$ cm$^{-2}$ at the radius of 1\farcs6 = 126 pc (the average half width of the synthesized beam). This gives $\xi$ of 1.3$\times10^{-15}$ s$^{-1}$.  We have also estimated that $n_{\rm H_{2}}$ of the nucleus is $10^{4\pm1}$ cm$^{-3}$ \citep{hsieh08}.  Hence $\log \xi$/$n_{\rm H_{2}}$ is --17.9 to --19.9.
For the other case of the $N_{\rm H_2}$ estimated from the X-ray
absorption, the corresponding $\xi$ is $10^{-13}$ s$^{-1}$
and the $\log \xi$/$n_{\rm H_{2}}$ is --15.9 to --17.9.
The HCN fractional abundance predicted
by the model corresponds to the range estimated by CO
and X-ray absorption are --8.7 to --7.4 and $\le$--7.8, respectively.
Thus, the HCN abundance predicted by the X-ray ionization
chemistry is consistent with both cases we estimated above.
However, if it favors the latter case, a larger H$_{2}$ number
density is probably required to generate the higher HCN abundance.
We also find that the HCO$^{+}$
abundance is consistent with the X-ray ionization chemistry
for the case of $N_{\rm H_2}$ as derived from X-ray absorption.
We conclude from our J = 3--2
high resolution studies, that the X-ray ionization
chemistry can explain the physical conditions
in the nucleus.

\section{SUMMARY}

\begin{enumerate}

\item We detected the J = 3--2 HCN, HCO$^{+}$
lines in the circumnuclear region of NGC 1097.
The HCN and HCO$^{+}$ lines reveal
a bright nucleus as compared to the molecular ring. The HCN(J = 3--2)
lines contribute $\sim$30\% of the total flux to the nucleus,
while it is only $\sim$20\% for the HCO$^{+}$(J = 3--2)
and CO(J =3--2) lines. The HCN(J = 3--2)/HCO$^{+}$(J = 3--2)
has a value of 2 in the nucleus and unity in the ring.
The J = 3--2 to J = 1--0 ratios of both lines are
as low as $\sim$0.2 for both the nucleus and the ring.
Our results generally agree with the previous single dish surveys.

\item Our standard LTE analysis suggests that
the HCN(J = 3--2) and HCO$^{+}$(J = 3--2)
lines might be optically thin in the nucleus. Both
lines can be optically thin or thick in the molecular
ring. The high-J lines with larger critical density,
however, might also be subthermally excited.
The tight correlation of the flux of HCN and HCO$^{+}$
and 24$\micron$ emission suggest these two lines trace
the same origins of star formation in the ring better than CO, while
other mechanisms such as AGN activities contribute
to the HCN enhancements in the nucleus.

\item The simple estimates of the self-consistency
of the X-ray ionization chemistry is made
with our high-J observations. The results suggest
that the ionization chemistry model can
explain the excess of the HCN abundance in the nucleus.

\end{enumerate}

\acknowledgements
We thank for the SMA staffs for the SMA operations
for this project. We acknowledges the anonymous
referee for the stimulating comments to improve the manuscript.
P.-Y. Hsieh acknowledges Dr. Tatsuhiko Hasegawa to
complete this project.
%This project was founded by the NSC...

\clearpage

%-------High resolution ratios

\begin{deluxetable}{cccccccc}
\tablewidth{0pt}
\tabletypesize{\scriptsize}
\setlength{\tabcolsep}{0.02in}
\tablecaption{Ratios at high angular resolution \label{t.high.ratio}}
\tablehead{\colhead{GMAs} & \colhead{Positions} & 
\colhead{HCN(J = 3--2)} &
\colhead{HCO$^{+}$(J = 3--2)} &
\colhead{$^{12}$CO(J = 3--2)} &
\colhead{HCN$_{32}$/HCO$^{+}$$_{32}$} &
\colhead{HCN$_{32}$/$^{12}$CO$_{32}$} &
\colhead{HCO$^{+}$$_{32}$/$^{12}$CO$_{32}$}\\
\colhead{} & \colhead{}  & \colhead{(1)} & \colhead{(2)} & \colhead{(3)} &
\colhead{(4)} &\colhead{(5)} & \colhead{(6)} \\
\colhead{} & \colhead{}  & \colhead{(K km s$^{-1}$)} & \colhead{(K km s$^{-1}$)} & \colhead{(K km s$^{-1}$)} & \colhead{} & \colhead{} & \colhead{}
}

\startdata
1 & 3\farcs0,9\farcs5	&	10.8	&	8.1	&	189.0	&	1.33$\pm$0.34	&	0.05$\pm$0.01	&	0.05$\pm$0.01	\\
2 & --5\farcs0,7\farcs5 &	7.6	&	5.2	&	120.9	&	1.47$\pm$0.57	&	0.06$\pm$0.01	&	0.05$\pm$0.01	\\
3 & --8\farcs5,--3\farcs0 &	9.5	&	8.5	&	273.4	&	1.11$\pm$0.29	&	0.03$\pm$0.01	&	0.04$\pm$0.01	\\
4 & --5\farcs0,--8\farcs0 &	7.9	&	7.5	&	159.5	&	1.06$\pm$0.32	&	0.04$\pm$0.01	&	0.04$\pm$0.01	\\
5 & 1\farcs0,--9\farcs0 &	6.4	&	4.8	&	110.4	&	1.36$\pm$0.59	&	0.06$\pm$0.02	&	0.06$\pm$0.02	\\
6 & 8\farcs5,--6\farcs5 &	6.0	&	9.7	&	121.9	&	0.62$\pm$0.20	&	0.04$\pm$0.01	&	0.06$\pm$0.01	\\
7 & 8\farcs5,4\farcs0 &	8.1	&	9.7	&	175.3	&	0.84$\pm$0.22	&	0.04$\pm$0.01	&	0.05$\pm$0.01	\\
8 & 0\farcs5,0\farcs0 &	24.7 &	11.6 &	336.1	&	2.13$\pm$0.34	&	0.07$\pm$0.01	&	0.04$\pm$0.01	
\enddata

\tablecomments{
The flux density of the peaks are measured within one synthesized
beam (4\farcs4$\times$2\farcs7). The ($\delta$R.A.,$\delta$Decl.) position of the peaks relative to the phase center. The uncertainties of
the HCN(J = 3--2), HCO$^{+}$(J = 3--2), and CO(J = 3--2)
are 1.7 K km s$^{-1}$, 1.7 K km s$^{-1}$, and 3.6 K km s$^{-1}$,
respectively.}
\end{deluxetable}

%------Low resolution ratios

\begin{deluxetable}{lcc}
\tablewidth{0pt}
\tabletypesize{\scriptsize}
\setlength{\tabcolsep}{0.02in}
\tablecaption{Ratios at low angular resolution \label{t.low.ratio}}
\tablehead{\colhead{Ratios} & 
\colhead{Nucleus} &
\colhead{Ring} \\
\colhead{} & \colhead{(1)} & \colhead{(2)}
}

\startdata
HCN$_{10}$/HCO$^{+}$$_{10}$	& 2.22$\pm$0.50  &	1.67$\pm$0.83	\\
HCN$_{10}$/CO$_{10}$			 	&	0.59$\pm$0.13 &	0.35$\pm$0.14\\
HCO$^{+}$$_{10}$/CO$_{10}$		&	0.27$\pm$0.07 &	0.21$\pm$0.11\\ \hline
HCN$_{32}$/HCO$^{+}$$_{32}$	&	2.01$\pm$0.41 &	1.07$\pm$0.48\\
HCN$_{32}$/CO$_{32}$	 			&	0.04$\pm$0.01 & 0.02$\pm$0.01\\
HCO$^{+}$$_{32}$/CO$_{32}$ 	&	0.02$\pm$0.01 &	0.02$\pm$0.01\\ \hline
HCN$_{32}$/HCN$_{10}$			&	0.15$\pm$0.02	&	0.12$\pm$0.05\\
HCO$^{+}$$_{32}$/HCO$^{+}$$_{10}$	&	0.17$\pm$0.05 & 0.19$\pm$0.10\\
$^{12}$CO$_{32}$/$^{12}$CO$_{10}$ &	2.33$\pm$0.44 & 1.84$\pm$0.56
\enddata

\tablecomments{
In this table, we show the low resolution (7\farcs9$\times$3\farcs2)
intensity ratios of the nucleus and ring. The ratio of the nucleus
is measured at the intensity peak. We measured the azimuthally
averaged intensity ratios of the ring within radius of 6\arcsec--13\arcsec. 
}
\end{deluxetable}

%------6" ratios and 24 micron

\begin{deluxetable}{ccccc}
\tablewidth{0pt}
\tabletypesize{\scriptsize}
\setlength{\tabcolsep}{0.02in}
\tablecaption{Comparisons of flux of 24\micron~and molecular lines \label{t.24.ratio}}
\tablehead{\colhead{Peaks} & 
\colhead{HCN(J = 3--2)} &
\colhead{HCO$^{+}$(J = 3--2)} &
\colhead{$^{12}$CO(J = 3--2)} &
\colhead{24\micron} \\
\colhead{} & \colhead{(K km s$^{-1}$)} & \colhead{(K km s$^{-1}$)} & \colhead{(K km s$^{-1}$)} & \colhead{(mJy $\rm arcsec^{-2}$)}
}

\startdata
1 &	5.0$\pm$0.8 & 5.5$\pm$0.8 & 100.2$\pm$1.8 & 114.49$\pm1.64$\\
2 &	4.1$\pm$0.8 & 3.6$\pm$0.8 & 72.2$\pm$1.8 & 113.82$\pm1.64$\\
3 &	4.3$\pm$0.8 & 4.4$\pm$0.8	& 146.7$\pm$1.8 &	102.97$\pm1.56$\\
4 &	3.9$\pm$0.8 & 4.7$\pm$0.8	& 100.4$\pm$1.8 &	82.95$\pm1.40$\\
5 &	4.0$\pm$0.8 & 3.1$\pm$0.8 & 75.7$\pm$1.8 & 77.99$\pm1.35$\\
6 & 	3.2$\pm$0.8 & 2.9$\pm$0.8 & 72.4$\pm$1.8 & 56.66$\pm1.15$\\
7 &	3.9$\pm$0.8 & 5.4$\pm$0.8	& 99.0$\pm$1.8 & 85.99$\pm1.42$\\
8 &	12.3$\pm$0.8 & 5.6$\pm$0.8  & 190.1$\pm$1.8 & 68.91$\pm1.27$
\enddata

\tablecomments{
We show the low resolution (6\arcsec)
intensity of the nucleus and ring, as well as the
Spitzer 24$\micron$ continuum emission. The unit of
the flux density of 24$\micron$~ continuum emission
is mJy $\rm arcsec^{-2}$ (10$^{-3}$ erg s$^{-1}$ cm$^{-1}$ $\rm arcsec^{-2}$
$\rm Hz^{-1}$) integrated over 6\arcsec.
}
\end{deluxetable}

\clearpage

\end{document}